\documentclass{iau}

\usepackage{amsmath}
\usepackage{graphicx}
\usepackage{multirow} 

\begin{document}

\lefttitle{Wladimir E. Banda-Barrag\'an et al.}
\righttitle{A cloud-cloud collision in Sgr B2?}

\jnlPage{1}{7}
\jnlDoiYr{2021}
\doival{10.1017/xxxxx}

\aopheadtitle{IAU Symposium No. 362}
\editors{Dmitry Bisikalo, Tomoyuki Hanawa, Christian Boily, eds.}

\title{A cloud-cloud collision in Sgr B2?\\ 3D simulations meet SiO observations}

\author{Wladimir E. Banda-Barrag\'an$^1$, Jairo Armijos-Abenda\~no$^2$, and Helga D\'enes$^3$}

\affiliation{$^1$Hamburger Sternwarte, Universit\"{a}t Hamburg, Gojenbergsweg 112, D-21029 Hamburg, Germany \\ email: {\tt wbanda@hs.uni-hamburg.de} \\
$^{2}$Observatorio Astron\'omico de Quito, Escuela Polit\'ecnica Nacional, Interior del Parque La Alameda, 170136, Quito, Ecuador\\
$^{3}$ASTRON - The Netherlands Institute for Radio Astronomy, NL-7991 PD Dwingeloo, The Netherlands
} 

\begin{abstract}
We compare the properties of shocked gas in Sgr B2 with maps obtained from 3D simulations of a collision between two fractal clouds. In agreement with $^{13}$CO(1-0) observations, our simulations show that a cloud-cloud collision produces a region with a highly turbulent density substructure with an average $N_{\rm H2}\gtrsim 5\times10^{22}\,\rm cm^{-2}$. Similarly, our numerical multi-channel shock study shows that colliding clouds are efficient at producing internal shocks with velocities of $5-50\,\rm km\,s^{-1}$ and Mach numbers of $\sim4-40$, which are needed to explain the $\sim 10^{-9}$ SiO abundances inferred from our SiO(2-1) IRAM observations of Sgr B2. Overall, we find that both the density structure and the shocked gas morphology in Sgr B2 are consistent with a $\lesssim 0.5\,\rm Myr$-old cloud-cloud collision. High-velocity shocks are produced during the early stages of the collision and can ignite star formation, while moderate- and low-velocity shocks are important over longer time-scales and can explain the extended SiO emission in Sgr B2.
\end{abstract}


\maketitle

\section{Introduction}
The Galactic centre, at a distance of $\sim 8\,\rm kpc$ (\citealt{Boehle2016,2019A&A...625L..10G}), contains $\sim3\times10^7\,\rm M_{\odot}$ of molecular gas with most of it lying inside a ring of clouds known as the Central Molecular Zone (CMZ). As a result, this zone harbours several regions of active star formation. One of such regions is Sgr B2, which contains $\sim 10^6-10^7\,\rm M_{\odot}$ of molecular gas (\citealt{Molinari2011,2021A&A...649A..32S}). Sgr B2 is located at a projected distance of $\sim 100$ pc away from the very centre of the Galaxy, which hosts Sgr A*, a supermassive black hole with an estimated mass of $4\times 10^6\,\rm M_{\odot}$ (\citealt{2019A&A...625L..10G}). Given its location, the CMZ is subjected to strong tidal forces, which makes the local interstellar medium (ISM) highly pressurised when compared to the ISM in the disc of the Galaxy. Typical thermal pressures in the CMZ are $\sim 10^6-10^7\,\rm K\,cm^{-3}$ (see \citealt{1992Natur.357..665S,2021A&A...649A..32S}), which are $\sim 1-2$ dex higher (see \citealt{2012MNRAS.423.3512C}) than the typical values of $\sim 10^4\,\rm K\,cm^{-3}$ found in the Galactic disc.\par

Dynamical models of the CMZ propose that the cloud complexes in this region follow either a closed $\infty$-shaped orbit (see \citealt{Molinari2011}) or open orbits with at least four gas streams (see \citealt{2015MNRAS.447.1059K}). Sgr B2 is located at the Eastern end of the CMZ (\citealt{1995ApJ...451..284D}), at the intersection between the so-called x$_1$ and x$_2$ orbits of the inner Galaxy, which are produced by the gravitational effects of the Galactic bar (\citealt{1991MNRAS.252..210B}). In the Molinari et al. scenario, which is also supported by previous authors (e.g. \citealt{Hasegawa1994}), Sgr B2 has an orbital speed of $\sim 80\,\rm km\,s^{-1}$ and is the result of collisions of gas travelling along these two orbits. On the other hand, in the Kruijssen et al. scenario, Sgr B2 has an orbital speed of $\sim 130\,\rm km\,s^{-1}$ and the star formation observed in it is due to gas compression caused by this cloud having passed the pericentre of the Galaxy $\sim 0.7\,\rm Myr$ ago. In order to assess whether the cloud-cloud collision or the pericentre-passage scenarios or both can explain the formation and evolution of Sgr B2, combining information from observations and numerical simulations is essential.\par

In this report, we briefly summarise the findings of our recent paper, \cite{2020MNRAS.499.4918A}, in which we report and explain new observations of SiO emission in Sgr B2, together with new hydrodynamical simulations of cloud-cloud collisions. Since this report only provides a summary of our findings, we encourage the readers to read our full paper for a more thorough discussion.

\section{Shocked gas structure and kinematics in Sgr B2}
As in all the other molecular clouds in the CMZ, the kinematics of Sgr B2 is complex. Observations reveal that this region contains hot cores with number densities $>10^6\,\rm cm^{-3}$ embedded in a dense envelope with number densities $\sim 10^5\,\rm cm^{-3}$ covering a few parsecs, and a more diffuse envelope with number densities $\sim 10^3\,\rm cm^{-3}$ covering a diameter of $\sim 40\,\rm pc$ (\citealt{Schmiedeke2016}). In this paper, we study the large-scale structure of Sgr B2 on scales covering ($15'\times15'$), equivalent to an area of $(36\times36)\,\rm pc^2$ (at the distance of the Galactic centre).\par

Using the IRAM 30-m telescope, we have observed Sgr B2 and detected several molecular lines in emission, including SiO J=2-1 at $86.8\,\rm GHz$, C$^{18}$O J=1-0 at $109.8\,\rm GHz$, and $^{13}$CO J=1-0 at $110.2\,\rm GHz$ (see the full sample in Table 1 of \citealt{2020MNRAS.499.4918A}). The first molecular line, SiO J=2-1, is a shock tracer (\citealt{1997A&A...321..293S,2016A&A...595A.122L}), which we use to study the structure and kinematics of shocked gas in this region, the second one, C$^{18}$O J=1-0, allows us to estimate the SiO abundances with respect to molecular hydrogen ($\rm H_{2}$), and the third molecular line, $^{13}$CO J=1-0, provides information on the typical hydrogen column densities. From the latter two, we find SiO relative abundances $N_{\rm SiO}/N_{\rm H_2}\sim 10^{-9}$, and a mean column number density of $\bar{N}_{\rm H_2}\gtrsim 5\times10^{22}\,\rm cm^{-2}$, respectively.\par

\begin{figure}[h!]
   \centering
   \includegraphics[width=0.95\textwidth]{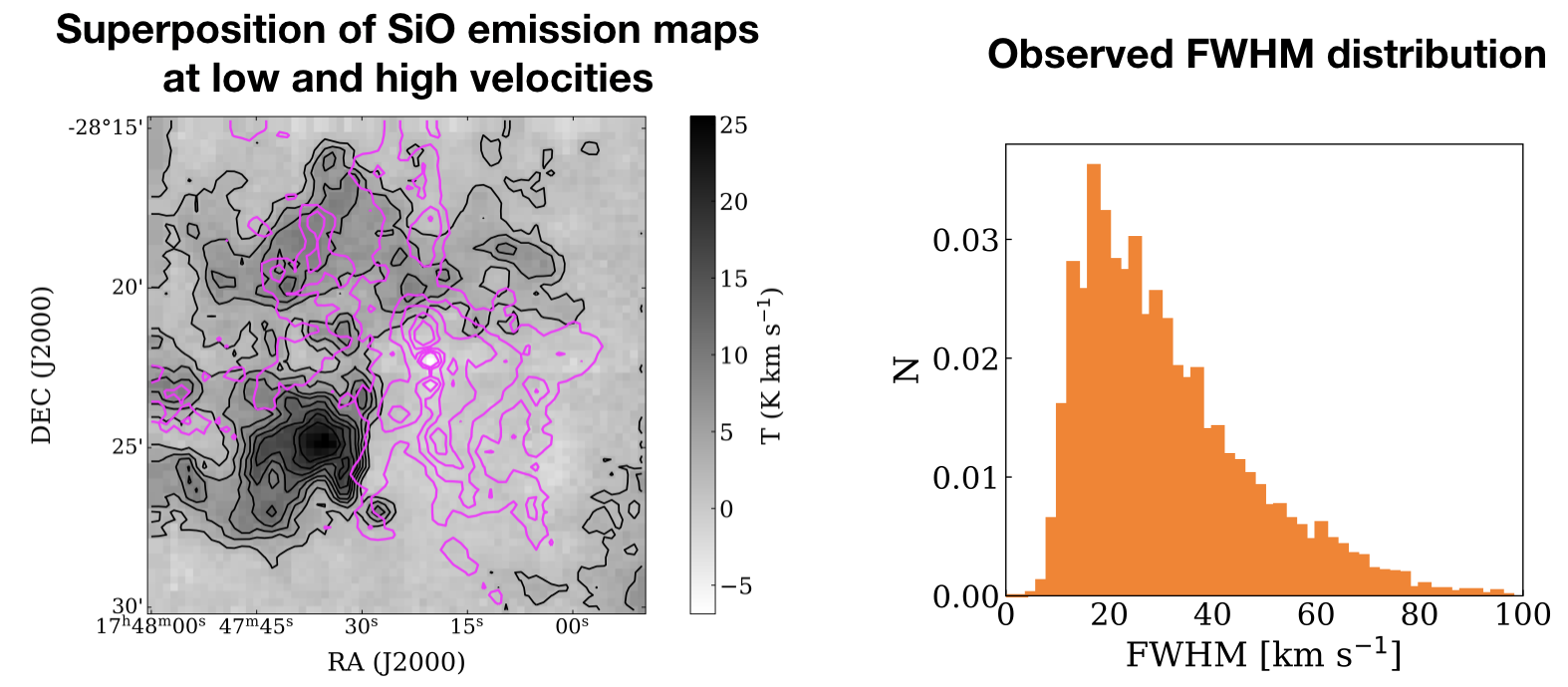}
   \caption{Left: Superimposed maps of SiO emission at low (black) and high (magenta) velocities, showing spatial anti-correlation. Black contour levels ([10, 25] km s$^{-1}$) are: 2.5 , 5, 7.5, 10, 12.5, 15, 17.5, 20 K km s$^{-1}$; magenta contour levels ([70, 85] km s$^{-1}$) are: 3, 6, 9, 12, 15 K km s$^{-1}$. Right: Distribution of the FWHM of SiO J=2-1 line emission. These panels have been adapted from \cite{2020MNRAS.499.4918A}, the reader is referred to that paper for further details.}
   \label{Fig1}
\end{figure}

We find that the SiO emission is very extended in Sgr B2, covering the full $(36\times36)\,\rm pc^2$ surveyed area and displaying a turbulent substructure with several arcs, cavities, and cores. Our observations also unveil a complex shocked gas kinematics as SiO emission covers a wide range of velocities, $[-5,+115]\,\rm km\,s^{-1}$. The spatial distribution of gas in different velocity channels reveal important properties about shocked gas in this region. Our maps indicate that shocked gas is turbulent in all velocity channels, and spatially anti-correlated at low and high velocities. Figure \ref{Fig1} shows two sample maps of integrated SiO emission at high velocities $[70,85]\,\rm km\,s^{-1}$ and low velocities $[10,25]\,\rm km\,s^{-1}$. These maps are spatially complementary, in agreement with expectations from cloud-cloud collision scenarios proposed earlier by e.g. \cite{Sato00} based on a similar complementarity found based on $^{13}$CO maps.\par

This spatial complementarity could also be caused by the superposition of clouds disconnected in 3D along the line of sight, but our position-velocity maps (such as the one displayed in Figure \ref{Fig1}) show that the structures at low and high velocities are connected by bridges and V-shaped features, which are also characteristic of cloud-cloud collisions (see \citealt{2017ApJ...835..142T,Enokiya19}). Therefore, we find that these structures are kinematically connected and likely belong to parcels of gas from the colliding clouds or gas streams, which would have been moving at different initial speeds. Similarly, we find that stellar feedback is unlikely to create the large-scale structure of Sgr B2 as we find no clear connection between broad SiO components and the local star-forming regions as expected in regions where stellar feedback plays a more important role (\citealt{2010MNRAS.406..187J}). The small sizes $<0.3\,\rm pc$ of the local H II regions (\citealt{Mehringer93}) also suggests stellar feedback would only be important in shaping sub-pc to pc structures, and that star formation, rather than the cause, is a product of the strong supersonic turbulence driven by e.g. the collision of gas clouds.\par

\section{A cloud-cloud collision in Sgr B2}
Cloud-cloud collisions are ubiquitous in ISM interactions (e.g. in shock-multicloud models, see \citealt{2020MNRAS.499.2173B,2021MNRAS.506.5658B}), but can a cloud-cloud collision really explain the properties of shocked gas in Sgr B2? To answer this question and confirm whether or not a cloud-cloud collision can produce Sgr B2 and its star formation, we carry out two analyses: 1) we decompose the observed SiO spectra using GaussPy+ \citep{Riener_2019} in order to study the typical SiO integrated line intensities and FWHM line widths in this region, and 2) we compare these line widths, which we use as shock velocity proxies (in fact we assume that $v_{\rm shock}\lesssim$ FWHM), to shock speeds measured directly from numerical simulations of cloud-cloud collisions (see Section 7.2 of \citealt{2020MNRAS.499.4918A}).\par

From the first analysis, we find SiO integrated line intensities of the order of 11 $\pm$ 3 K km s$^{-1}$, peaking at $\sim$ 5 K km s$^{-1}$, and SiO line widths of 31 $\pm$ 5 km s$^{-1}$, slightly higher than the background gas $^{13}$CO, peaking at 21 km s$^{-1}$. The fact that SiO emission line widths are in the range of $5-50\,\rm km\,s^{-1}$ suggests that shocks in Sgr B2 are predominantly moderate- and low-velocity shocks. Since we do not find a correlation between star-forming regions and zones with broad SiO lines, we argue that shocks in Sgr B2 emerge in supersonically-turbulent gas that has been produced by stirring after a cloud-cloud collision.\par

\begin{figure}[h!]
   \centering
   \includegraphics[width=0.95\textwidth]{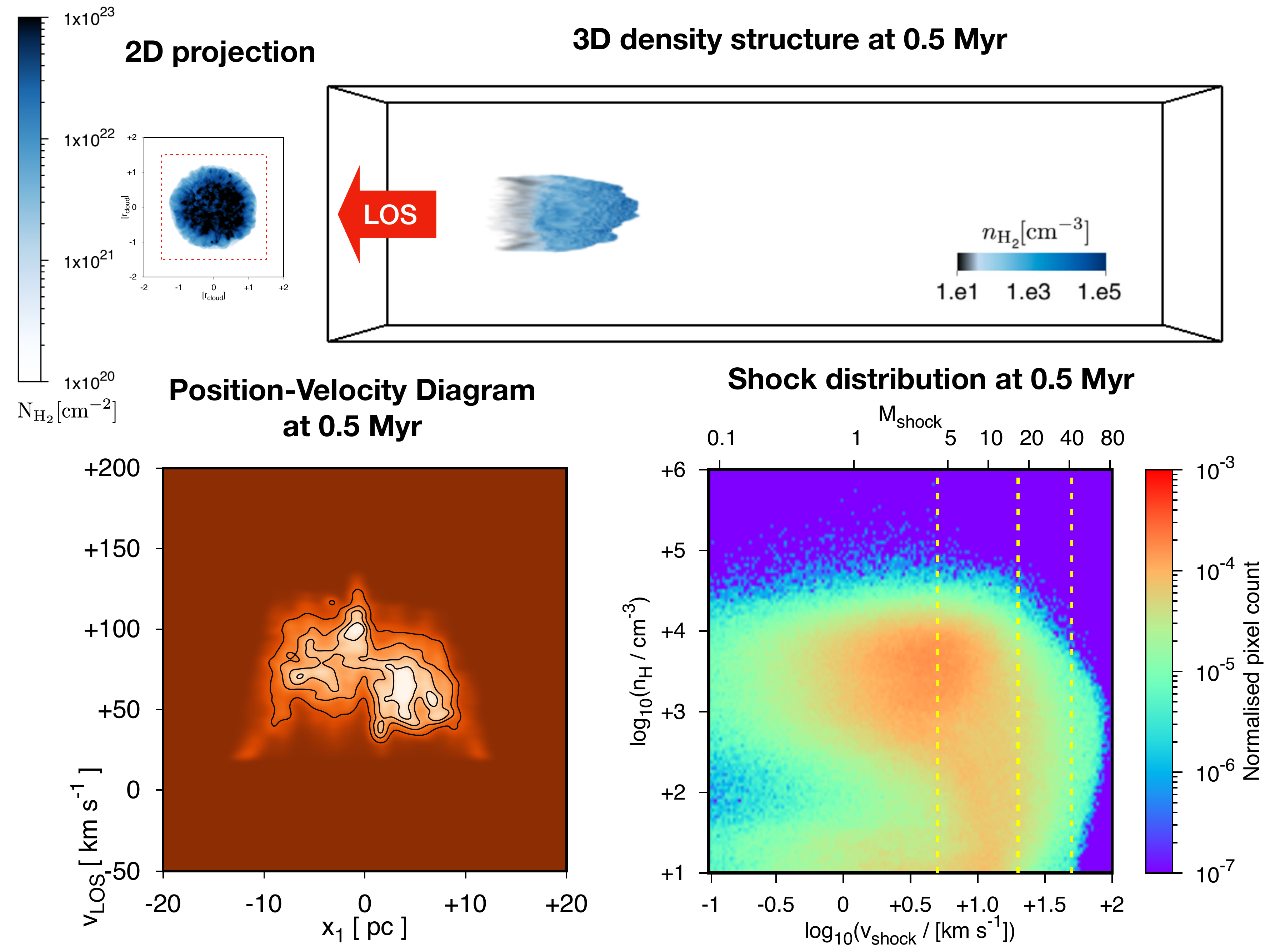}
   \caption{Cloud-cloud collision simulation at $0.5\,\rm Myr$ showing the 3D density structure (top right corner), the 2D column numbers density projection (top left corner), and the corresponding position-velocity diagram and shock distribution at the same time. These panels have been adapted from \cite{2020MNRAS.499.4918A}, the reader is referred to that paper and the simulation movies at \url{https://tinyurl.com/y5bc3smn} for further details.}
   \label{Fig2}
\end{figure}

From the second analysis regarding numerical simulations (see Figure \ref{Fig2}), we find that colliding fractal clouds are efficient at producing internal shocks with velocities $\sim 5-50\,\rm km\,s^{-1}$ and typical shock Mach numbers of $\sim4-40$ (akin to those reported by \citealt{2016MNRAS.457.2675H} for the CMZ). In our models, clouds are initially moving at a relative velocity of $120\,\rm km\,s^{-1}$ as we find that speed is needed to produce shocks in high-velocity channels. Similarly, our simulations indicate that shocked gas (i.e. SiO emission) is efficiently produced during the early stages of the collision for $t<0.5\,\rm Myr$ in all velocity channels. Most of the emission is concentrated at velocities in the range of $[25,70]\,\rm km\,s^{-1}$. As gas decelerates following the collision, we find that shocked gas with lower and higher velocities than those limits quickly disappears at later stages of the evolution, allowing us to constrain the age of the collision. Position-velocity diagrams for SiO, obtained from the simulations, also show that V-shaped structures are short-lived and disappear at later stages, which further confirms the $t<0.5\,\rm Myr$ time-scale. This time-scale is consistent with the estimated age of the shell-like structures in Sgr B2 reported by \cite{Tsuboi15}.\par

During the collision we find that the distribution of shocks also varies with time. At the very early stages of the collision for $t<0.2\,\rm Myr$, the dominant shocks are high-velocity shocks with $v_{\rm shock}>50\,\rm km\,s^{-1}$, which can easily ignite star formation on time-scales of $\sim 0.1\,\rm Myr$. Between $t=0.2\,\rm Myr$ and $0.5\,\rm Myr$, the dominant shocks are moderate- and low-velocity shocks with speeds in the range of $v_{\rm shock}=5-50\,\rm km\,s^{-1}$, which can maintain the widespread SiO emission. This range of shock velocities is also consistent with other shock models that reproduce the observed mid-$J$ CO emission in Sgr B2 (see \citealt{2021A&A...649A..32S}). On the other hand, for $t>0.5\,\rm Myr$, very-low-velocity shocks with $v_{\rm shock}=2-5\,\rm km\,s^{-1}$ and subsonic waves dominate the shock distribution. Thus, in this scenario the source of SiO emission is replenishment by a population of moderate- and low-velocity shocks as they have the appropriate velocities to trigger grain mantle sputtering (\citealt{Gusdorf08a}), which generally needs shocks speeds $>7\, \rm km\, s^{-1}$, within the reasonably-long timescales of $\lesssim 10^5\,\rm yr$ (\citealt{Harada15}), which are needed to explain the observed SiO abundance in this region.

\section{Conclusion}
Our IRAM observations of SiO J=2-1 emission in Sgr B2 show that shocked gas in this region is widespread and has a complex kinematics covering a wide velocity range, $[-5,+115]\,\rm km\,s^{-1}$. Shocked gas in this region has a turbulent substructure with a fractal morphology characterised by arcs, cavities, and cores. The spatial anti-correlation of gas at low and high velocities, and the presence of V-shaped features on position-velocity maps strongly suggests that Sgr B2 is a product of a cloud-cloud collision between gas likely travelling along the x$_1$ and x$_2$ orbits of the inner Galaxy.\par

Our numerical simulations of collisions between fractal clouds suggest that a cloud-cloud collision that took place $\lesssim 0.5\,\rm Myr$ ago can readily explain both the gas density structure and the distribution and kinematics of shocked gas in Sgr B2. During the early stages of the collision, high-velocity shocks are produced with the ability to ignite star formation. Later on, shocks evolve into moderate- and low-velocity shocks with speeds in the range of $5$ to $50\,\rm km\,s^{-1}$, which can explain the widespread SiO emission. Thus, in this scenario, Sgr B2 can be interpreted as a structure produced by turbulent stirring associated with colliding clouds or colliding gas streams.\par

The authors gratefully acknowledge the Gauss Centre for Supercomputing e.V. (\url{www.gauss-centre.eu}) for funding this project by providing computing time (via grant pn34qu) on the GCS Supercomputer SuperMUC-NG at the Leibniz Supercomputing Centre (\url{www.lrz.de}). WEBB is supported by the Deutsche Forschungsgemeinschaft (DFG) via grant BR2026/25, and by the National Secretariat of Higher Education, Science, Technology, and Innovation of Ecuador, SENESCYT, via grant 1711298438. We also thank the referee for their very helpful comments and suggestions, which helped us improve this paper.

\bibliographystyle{iaulike}
\bibliography{bibliography}

\begin{thebibliography}{}

\bibitem[{Armijos-Abenda{\~n}o} et~al., 2020]{2020MNRAS.499.4918A}
{Armijos-Abenda{\~n}o}, J., {Banda-Barrag{\'a}n}, W.~E., {Mart{\'\i}n-Pintado},
  J., {D{\'e}nes}, H., {Federrath}, C., \& {Requena-Torres}, M.~A. 2020,
  {Structure and kinematics of shocked gas in Sgr B2: further evidence of a
  cloud-cloud collision from SiO emission maps}.
\newblock {\em \mnras}, 499(4), 4918--4939.

\bibitem[{Banda-Barrag{\'a}n} et~al., 2020]{2020MNRAS.499.2173B}
{Banda-Barrag{\'a}n}, W.~E., {Br{\"u}ggen}, M., {Federrath}, C., {Wagner},
  A.~Y., {Scannapieco}, E., \& {Cottle}, J. 2020, {Shock-multicloud
  interactions in galactic outflows - I. Cloud layers with lognormal density
  distributions}.
\newblock {\em \mnras}, 499(2), 2173--2195.

\bibitem[{Banda-Barrag{\'a}n} et~al., 2021]{2021MNRAS.506.5658B}
{Banda-Barrag{\'a}n}, W.~E., {Br{\"u}ggen}, M., {Heesen}, V., {Scannapieco},
  E., {Cottle}, J., {Federrath}, C., \& {Wagner}, A.~Y. 2021, {Shock-multicloud
  interactions in galactic outflows - II. Radiative fractal clouds and cold gas
  thermodynamics}.
\newblock {\em \mnras}, 506(4), 5658--5680.

\bibitem[{Binney} et~al., 1991]{1991MNRAS.252..210B}
{Binney}, J., {Gerhard}, O.~E., {Stark}, A.~A., {Bally}, J., \& {Uchida}, K.~I.
  1991, {Understanding the kinematics of Galactic Centre gas.}
\newblock {\em \mnras}, 252, 210.

\bibitem[{Boehle} et~al., 2016]{Boehle2016}
{Boehle}, A., {Ghez}, A.~M., {Sch\"odel}, R., {Meyer}, L., {Yelda}, S.,
  {Albers}, S., {Martinez}, G.~F., {Becklin}, E.~E., \& {Do}, T. 2016, {}.
\newblock {\em \apj}, 830, 17.

\bibitem[{Crocker}, 2012]{2012MNRAS.423.3512C}
{Crocker}, R.~M. 2012, {Non-thermal insights on mass and energy flows through
  the Galactic Centre and into the Fermi bubbles}.
\newblock {\em \mnras}, 423(4), 3512--3539.

\bibitem[{de Pree} et~al., 1995]{1995ApJ...451..284D}
{de Pree}, C.~G., {Gaume}, R.~A., {Goss}, W.~M., \& {Claussen}, M.~J. 1995,
  {The Sagittarius B2 Star-forming Region. II. High-Resolution H66 alpha
  Observations of Sagittarius B2 North}.
\newblock {\em \apj}, 451, 284.

\bibitem[{Enokiya} et~al., 2019]{Enokiya19}
{Enokiya}, R., {Torii}, K., \& {Fukui}, Y. 2019, {}.
\newblock {\em \pasj}, 00, 1--16.

\bibitem[{Gravity Collaboration} et~al., 2019]{2019A&A...625L..10G}
{Gravity Collaboration}, {Abuter}, R., {Amorim}, A., {Baub{\"o}ck}, M.,
  {Berger}, J.~P., {Bonnet}, H., {Brandner}, W., {Cl{\'e}net}, Y., {Coud{\'e}
  Du Foresto}, V., {de Zeeuw}, P.~T., {Dexter}, J., {Duvert}, G., {Eckart}, A.,
  {Eisenhauer}, F., {F{\"o}rster Schreiber}, N.~M., {Garcia}, P., {Gao}, F.,
  {Gendron}, E., {Genzel}, R., {Gerhard}, O., {Gillessen}, S., {Habibi}, M.,
  {Haubois}, X., {Henning}, T., {Hippler}, S., {Horrobin}, M.,
  {Jim{\'e}nez-Rosales}, A., {Jocou}, L., {Kervella}, P., {Lacour}, S.,
  {Lapeyr{\`e}re}, V., {Le Bouquin}, J.~B., {L{\'e}na}, P., {Ott}, T.,
  {Paumard}, T., {Perraut}, K., {Perrin}, G., {Pfuhl}, O., {Rabien}, S.,
  {Rodriguez Coira}, G., {Rousset}, G., {Scheithauer}, S., {Sternberg}, A.,
  {Straub}, O., {Straubmeier}, C., {Sturm}, E., {Tacconi}, L.~J., {Vincent},
  F., {von Fellenberg}, S., {Waisberg}, I., {Widmann}, F., {Wieprecht}, E.,
  {Wiezorrek}, E., {Woillez}, J., \& {Yazici}, S. 2019, {A geometric distance
  measurement to the Galactic center black hole with 0.3\% uncertainty}.
\newblock {\em \aap}, 625, L10.

\bibitem[{Gusdorf} et~al., 2008]{Gusdorf08a}
{Gusdorf}, A., {Pineau des For\^ests}, G., {Cabrit}, S., \& {Flower}, D.~R.
  2008, {}.
\newblock {\em \aap}, 490, 695--706.

\bibitem[{Harada} et~al., 2015]{Harada15}
{Harada}, N., {Riquelme}, D., {Viti}, S., {Jim\'enez-Serra}, I.,
  {Requena-Torres}, M.~A., {Menten}, K.~M., {Mart\'in}, S., {Aladro}, R.,
  {Mart\'in-Pintado}, J., \& {Hochg\"urtel}, S. 2015, {}.
\newblock {\em \aap}, 584, A102.

\bibitem[{Hasegawa} et~al., 1994]{Hasegawa1994}
{Hasegawa}, T., {Sato}, F., {Whiteoak}, J.~B., \& {Miyawaki}, R. 1994, {}.
\newblock {\em \apj}, 429, L77--L80.

\bibitem[{Henshaw} et~al., 2016]{2016MNRAS.457.2675H}
{Henshaw}, J.~D., {Longmore}, S.~N., {Kruijssen}, J.~M.~D., {Davies}, B.,
  {Bally}, J., {Barnes}, A., {Battersby}, C., {Burton}, M., {Cunningham},
  M.~R., {Dale}, J.~E., {Ginsburg}, A., {Immer}, K., {Jones}, P.~A., {Kendrew},
  S., {Mills}, E.~A.~C., {Molinari}, S., {Moore}, T.~J.~T., {Ott}, J.,
  {Pillai}, T., {Rathborne}, J., {Schilke}, P., {Schmiedeke}, A., {Testi}, L.,
  {Walker}, D., {Walsh}, A., \& {Zhang}, Q. 2016, {Molecular gas kinematics
  within the central 250 pc of the Milky Way}.
\newblock {\em \mnras}, 457(3), 2675--2702.

\bibitem[{Jim{\'e}nez-Serra} et~al., 2010]{2010MNRAS.406..187J}
{Jim{\'e}nez-Serra}, I., {Caselli}, P., {Tan}, J.~C., {Hernand ez}, A.~K.,
  {Fontani}, F., {Butler}, M.~J., \& {van Loo}, S. 2010, {Parsec-scale SiO
  emission in an infrared dark cloud}.
\newblock {\em \mnras}, 406(1), 187--196.

\bibitem[{Kruijssen} et~al., 2015]{2015MNRAS.447.1059K}
{Kruijssen}, J.~M.~D., {Dale}, J.~E., \& {Longmore}, S.~N. 2015, {The dynamical
  evolution of molecular clouds near the Galactic Centre - I. Orbital structure
  and evolutionary timeline}.
\newblock {\em \mnras}, 447(2), 1059--1079.

\bibitem[{Louvet} et~al., 2016]{2016A&A...595A.122L}
{Louvet}, F., {Motte}, F., {Gusdorf}, A., {Nguy{\^e}n Luong}, Q., {Lesaffre},
  P., {Duarte-Cabral}, A., {Maury}, A., {Schneider}, N., {Hill}, T., {Schilke},
  P., \& {Gueth}, F. 2016, {Tracing extended low-velocity shocks through SiO
  emission. Case study of the W43-MM1 ridge}.
\newblock {\em \aap}, 595, A122.

\bibitem[{Mehringer} et~al., 1993]{Mehringer93}
{Mehringer}, D.~M., {Palmer}, P., {Goss}, W.~M., \& {Yusef-Zadeh}, F. 1993, {}.
\newblock {\em \apj}, 412, 684--695.

\bibitem[{Molinari} et~al., 2011]{Molinari2011}
{Molinari}, S., {Bally}, J., {Noriega-Crespo}, A., {Compi\`egne}, M.,
  {Bernard}, J.~P., {Paradis}, D., {Martin}, P., {Testi}, L., \& {Barlow}, M.
  2011, {}.
\newblock {\em \apjl}, 735, L33.

\bibitem[{Riener} et~al., 2019]{Riener_2019}
{Riener}, M., {Kainulainen}, J., {Henshaw}, J.~D., {Orkisz}, J.~H., {Murray},
  C.~E., \& {Beuther}, H. 2019, {GAUSSPY+: A fully automated Gaussian
  decomposition package for emission line spectra}.
\newblock {\em \aap}, 628, A78.

\bibitem[{Santa-Maria} et~al., 2021]{2021A&A...649A..32S}
{Santa-Maria}, M.~G., {Goicoechea}, J.~R., {Etxaluze}, M., {Cernicharo}, J., \&
  {Cuadrado}, S. 2021, {Submillimeter imaging of the Galactic Center starburst
  Sgr B2. Warm molecular, atomic, and ionized gas far from massive star-forming
  cores}.
\newblock {\em \aap}, 649, A32.

\bibitem[{Sato} et~al., 2000]{Sato00}
{Sato}, F., {Hasegawa}, T., {Whiteoak}, J.~B., \& {Miyawaki}, R. 2000, {}.
\newblock {\em \apj}, 535, 857--868.

\bibitem[{Schilke} et~al., 1997]{1997A&A...321..293S}
{Schilke}, P., {Walmsley}, C.~M., {Pineau des Forets}, G., \& {Flower}, D.~R.
  1997, {SiO production in interstellar shocks.}
\newblock {\em \aap}, 321, 293--304.

\bibitem[{Schmiedeke} et~al., 2016]{Schmiedeke2016}
{Schmiedeke}, A., {Schilke}, P., {M\"oller}, T., {S\'anchez-Monge}, A.,
  {Bergin}, E., {Comito}, C., {Csengeri}, T., {Lis}, D.~C., \& {Molinari}, S.
  2016, {}.
\newblock {\em \aap}, 588, A143.

\bibitem[{Spergel} and {Blitz}, 1992]{1992Natur.357..665S}
{Spergel}, D.~N. \& {Blitz}, L. 1992, {Extreme gas pressures in the galactic
  bulge}.
\newblock {\em \nat}, 357(6380), 665--667.

\bibitem[{Torii} et~al., 2017]{2017ApJ...835..142T}
{Torii}, K., {Hattori}, Y., {Hasegawa}, K., {Ohama}, A., {Haworth}, T.~J.,
  {Shima}, K., {Habe}, A., {Tachihara}, K., {Mizuno}, N., {Onishi}, T.,
  {Mizuno}, A., \& {Fukui}, Y. 2017, {Triggered O Star Formation in M20 via
  Cloud-Cloud Collision: Comparisons between High-resolution CO Observations
  and Simulations}.
\newblock {\em \apj}, 835(2), 142.

\bibitem[{Tsuboi} et~al., 2015]{Tsuboi15}
{Tsuboi}, M., {Miyazaki}, A., \& {Uehara}, K. 2015, {}.
\newblock {\em \pasj}, 67, 90.

\end{thebibliography}

\vspace{-0.2cm}
\section*{Discussion}

\textbf{Question 1: How do you do this morphological comparison between the simulations and the observations? It sounds like a very complicated task to make sense out of it.}\\

Answer 1: We basically carry out two comparisons. One-to-one comparisons are always very hard, but what we did was to take the SiO emission maps from the observations in different velocity channels and study the shock distributions in our simulations within the same velocity channels. In that way we compared the shock velocities that we got from the simulations with the line widths of SiO emission from the observations, and we found that they are consistent if the time-scale of the collision evolution is $<0.5\,\rm Myr$. The second comparison involves position-velocity diagrams, which display zig-zag and V-shaped features that we can also reproduce with our simulations within a similar time-scale. Both are qualitative comparisons.\\

\textbf{Question 2: Would machine learning help here to do such a comparison in all dimensions at the same time?}\\

Answer 2: Possibly yes. We actually did incorporate a bit of machine learning in the analysis. The spectral decomposition of the observed emission lines  was done using GaussPy+, which uses machine learning for the spectral decomposition. Machine learning would also very likely help to do multi-dimensional analyses on the data in the future.\\

\textbf{Question 3: Did you compare your simulations/observations with other molecular line emissions, like CO, e.g. $^{13}$CO?}\\

Answer 3:  Yes, in the paper we also included analyses for C$^{18}$O and $^{13}$CO, which also show similar kinematics and in the latter case also a spatial anti-correlation at low- and high- velocities similar to that seen in SiO emission. We also decomposed the $^{13}$CO spectra and found that it has a similar FWHM distribution than SiO, but it is a little bit shifted towards lower velocities. In general, SiO has larger FWHM than CO because gas is supersonically turbulent.\\

\textbf{Question 4: How about the spatial distribution of CO? I guess you can reproduce the main features.}\\

Answer 4: Yes, we can reproduce the main (fractal-like) density features too.\\

\textbf{Question 5: It seems like with these models where you have two colliding clouds, you get these two-velocity features with the bridge between them. It has never been clear to me if this is a unique feature of cloud-cloud collisions although it is always interpreted as a sign of it. I guess with your models, you could actually do different kinds of models that are not necessarily cloud-cloud collisions and see if you get any similar looking features. Have you thought about this?}\\

Answer 5: I think these bridges are quite unique features of cloud-cloud collisions because if we would have just two clouds moving at different velocities, not colliding but disconnected along the line of sight, these features would be absent. In the paper we also discussed how the initial density distribution of the colliding clouds affect the density maps and for example if we assume that the clouds do not have this turbulent substructure, but they are just spherical clouds with uniform densities, then it would be harder to reproduce the observations.\\

\textbf{Question 6: I was thinking more of the idea of for example if you have a cloud that has been torn apart by shear then you might see a feature like that.}\\

Answer 6: I agree. I think shear could also cause a feature like the one seen in cloud-cloud collisions.\\

\textbf{Question 7: In the movie of the colliding clouds you showed, on the left one you see these streamers at the back of this cloud. What are they?}\\

Answer 7: The simulation is idealised and I set up one of the clouds at rest while the other one is moving on the grid. Because we have one cloud initially moving across the domain, this cloud is going to suffer from stripping by Kelvin-Helmholtz instabilities. Therefore, these streamers are sort of tails that form from mass stripping via dynamical instabilities. The simulation movies are available at \url{https://tinyurl.com/y5bc3smn}.

\end{document}